\documentstyle[12pt]{article}

\begin{document}
\begin{center}

{\Large \bf   A 1-$d$ Ising model for ripple formation}

\vskip 1.0cm
{\large \bf Nicolas Vandewalle$^{1,2}$ and Serge Galam$^2$}

\vskip 0.6cm $^1$ GRASP, Institut de Physique B5, Universit\'e de
Li\`ege,\\ B-4000 Li\`ege, Belgium.
\vskip 0.6cm $^2$ Laboratoire des
Milieux D\'esordonn\'es et H\'et\'erog\`enes, Tour 13, Case 86, \\ 4 place
Jussieu, 75252 Paris Cedex 05, France.

\end{center}

\vskip 1.5cm

\begin{abstract}

A 1-$d$ Ising model is shown to reproduce qualitatively the dynamics of
ripple formation. Saltation effect is imposed using a Kawasaki dynamics and
a pair interaction over some distance $\ell$. Within this model, the ripple
state turns out to be metastable in agreement with cellular automata
simulations as well as recent under water experiments. A dynamical phase
diagram is obtained. A mean-field solution of the problem is given in terms
of the ripple size. A mapping is then performed onto  a two-dimensional
$\,\ell\times\infty\,$ static problem.

\end{abstract}
\vskip 1.5cm


\newpage

\section{Introduction}

Ising spin models have been used to describe a tremendous spectrum of
physical and non-physical problems \cite{compaction,galam}. Among others,
they have proven useful to understand some aspect of granular media. In
particular compaction dynamics under tapping \cite{compaction,compaction2}.

Herein, we present a simple 1-$d$ Ising model to describe qualitative
features of ripple formation. Previous experimental and theoretical works
have underlined the primary role played by saltation in the emergence of
sand ripples and dunes \cite{betat,anderson,bouchaud,valance}. Saltation
means grain ejection and eolian transport over long distances. The hopping
and rolling of grains have generally led to travelling ripple structures
which in turn merge and grow.

In the following, an Ising spin description of the dynamics of ripples
formation is presented. A 1-$d$ Hamiltonain is suggested. It is studied
numerically with a Kawasaki dynamics. A mean-field treatment is also
performed. A dynamical phase diagram is obtained. Results are discussed and
compared with recent cellular automaton results \cite{ijmpc} and with
recent underwater experiments \cite{betat}.

\section{The model}

A natural way to map a one-dimensional granular landscape into a spin model
is to consider a 1-$d$ Ising model $\{\sigma_i=\pm 1\}$. Let the wind
blowing from left to right. At each site $i$, $\sigma_i=+1$ if the
local slope is positive (exposed to the wind). It is $\sigma_i=-1$ if the
local slope is negative (screened by the ripple crest and unaffected by the
wind). Successive slopes being then associated to the succession of spin
domains. Spin domains represent the ripple sides. A sketch of
such a mapping is shown in Figure 1.

In Figure 1, a grain is first shifted over a distance $\ell=7$ by
saltation. An avalanche then occurs to relaxe the surface. From the
magnetic viewpoint, both extraction and relaxation of one grain produce two
pairs of spin flips meaning that the whole dynamics goes by spin pairs.
Moreover, it proves that the dynamics does not modify the total
magnetization $m = \sum_i \sigma_i$. To embody these physical effects
charaterizing the saltation-avalanche mechanism, we consider the so-called
Kawasaki dynamics \cite{binder}. Having a 1-$d$ Ising system the
equilibrium total magnetization is zero. In addition, it will stays zero
since the Kawasaki dynamics preserves the total magnetization.

Avalanches which relaxe the granular surface can be viewed as small thermal
fluctuations along the ripple sides. In our spin mapping such a scheme
corresponds to a ferromagnetic nearest neigbor coupling $J_{nn}$ divided by
$k_BT$ where $T$ is the temperature and $k_B$ the Boltzman's constant.
Moreover, saltation occurs only on the faces exposed to the wind and leads
to a perturbation at the position $i+\ell$ where $\ell$ is the distance
overwhich the sand is transported by saltation. As for avalanches, we mimic
this effect by a ferromagnetic coupling $J_{s}$ which connects a site $i$
to a site $i+\ell$, divided by $k_BT$.

On this basis we consider the following dimensionless Hamiltonian
\begin{equation} E = - K_{nn} \sum_{i=-\infty}^{+\infty} \sigma_i
\sigma_{i+1} - K_{s} \sum_{i=-\infty}^{+\infty} \sigma_i \sigma_{i+\ell}
\delta_{\sigma_i,+1}  \ , \end{equation} where $\delta_{x,y}$ is the
Kronecker function, $K_{nn} \equiv \frac{J_{nn}}{k_BT}$ and $K_{s} \equiv
\frac{J_{s}}{k_BT}$. The Kronecker delta function in the second term
assumes that only faces exposed to the wind (spins +1 for a wind blowing
from left to right) play a role for saltation. These assumptions seem
reasonable and not of drastic consequences at our present level of
investigation.

\section{Numerical simulations}

Numerical simulations have been performed on a 1-$d$ system with periodic
boundary conditions, i.e. a loop. The initial spin loop configuration is
random with $m=0$. Thereafter, one positive spin and one negative spin are
randomly selected at each time step. The energy variation $\Delta E$ due to
the flipping of both selected spins is calculated. Periodic boundary
conditions are used in the energy calculation. The selected spins are then
flipped with some probability $p={\exp{\Delta E} \over 1 + \exp{\Delta E}}$
as in classical Monte Carlo methods \cite{binder}. The fact that two
opposite spins are chosen at each time step preserves the $m=0$ value of
the order parameter during the whole simulation, i.e. a Kawasaki dynamics.

Since the magnetization $m$ is zero and constant, a dynamical``order
parameter" can be defined with the domain wall density $\rho$, i.e. the
density of nearest neighboring antagonist pairs of spins. One should note
that \begin{equation} \rho = {\xi \over L} \end{equation} where $\xi$ is
the mean size of the spin domains. Loop lengths up to $L=10000$ spins have
been used. Our numerical results has been checked against variation in the
parameters ($K_{nn}$, $K_s$ and $\ell$). Finite-size effects as well as the
spin pattern time evolution have also been analyzed. it is worth noting
that $\ell$ and $L$ have been always chosen to be incommensurate in order
to avoid any artefact due to the periodic boundary conditions.

Figure 2 presents typical space-time patterns in which domain walls are
drawn for characterisitic lengths $L=144$ and $\ell=21$. Three different
sets of parameter values $K_{nn}$ and $K_s$ are shown. Depending on the
parameter values three different regimes are obtained. Either disorder
(Figure 2a), or order with domains of roughly the size $\ell/2$ (Figure
2b), or also order with larger domains (Figure 2c). Domains having a size
of the order $\ell/2$ are called thereby ``ripples". In parallel much
larger domains are called ``dunes" in reference to simulations in
\cite{ijmpc}.

Figure 3 contains the domain wall density $\rho$ as a function of time for
the three regimes of Figure 2. It is observed: (i) an erratic signal for
the disordered regime presenting huge fluctuations of $\rho$, (ii) an
apparently stable density $\rho \approx \ell/2$ after a short transient
regime and (iii) a low density regime ($\rho \approx 0$) decorated with
small fluctuations after some transient regime.

Figure 4 is the associated resulting dynamical phase diagram as function of
$(K_{nn},K_s)$ obtained numerically for $\ell > 1$. Three different
dynamical regimes are thus found. A region of the disorder regime is
represented in grey in the phase diagram. The boundary with the order
regime is found to be stable whatever the values of $\ell > 1$ and $L$.
Curves $\rho(K_{nn})$ are shown in Figure 5 for different values of $K_s$
where one can clearly see a crossover around $K_{nn} \approx 1$. At this
point, the domain size is equivalent to $\ell /2$. Moreover, this boundary
is stable in time. This result could be astonishing since finite-size
effects are expected. One should recall that $m=0$ corresponds to the
equilibrium state of a one-dimensional Ising model and thus fluctuations
and the domain wall evolution take place in the equilibrium state. Those
are mainly driven by the Ising interaction $K_{nn}$.

For the order regime (white region in Figure 4), two distinct regions
coexist: ripples (for $K_s >> K_{nn}$) and dunes (for $K_{nn} >> K_s$). The
dashed line in Figure 4 represents the boundary between these regimes.
However, this boundary moves slightly when the system size $L$ or the
saltation length $\ell$ is changed. In fact, the ripple state is
metastable. After very long times, ripples can suddenly merge into dunes
(as seen by geologists). The larger $K_s$, the more stable ripples are.

Different initial correlations have been examined. Starting from a perfect
antiferromagnetic situation, the system reaches the same regimes as
described above. The dynamics is similar to what we have observed for the
random inital situation. However, if one starts with a system containing
only two large domains, i.e. a single giant dune, the system never produces
ripples but remains unchanged in the order region of the dynamical phase
diagram. In the disorder regime (in grey in Figure 4), the initial giant
dune is however destroyed.

Finally, one should note the case $\ell=1$ is quite particular. In that
case, spin domains are produced even when $K_{nn} = 0$ for which a disorder
regime is expected. As soon as $\ell > 1$, the results summarized in Figure
4 are recovered.

\section{Mean-field treatment}

Eventhough the Hamiltonian (1) is rather simple, it does not allow for an easy
exact solution. For instance, the classical method of the transfer matrix
\cite{stanley} leads to searching for $2^{\ell+1} \times 2^{\ell+1}$
matrices. In order to solve this problem, we have considered as a first
approximation that the mean domain size $\xi$ is the relevant parameter.
Fluctuations in $\xi$ are neglected.

Assuming an order with domains of size $\xi$, the first product $\sigma_i
\sigma_{i+1}$ of Eq.(1) gives ${\xi-2 \over \xi}$ due to the presence of
domain walls which are energetically defavorable. The second product
$\sigma_i \sigma_{i+\ell}$ of Eq.(1) gives a periodically oscillating
function of period $2\xi$. Thus, the energy per lattice site $E/L$ can be
written as \begin{equation} {E \over L} \approx - K_{nn} {\left( {\xi-2
\over \xi} \right) } - {K_{s} \over 2} \cos{\left( {\pi \ell \over \xi}
\right)}  \ , \end{equation} which is drawn in Figure 6 as a function of
$\xi$ and for arbitrary values of the coupling energies $K_{nn}$ and
$K_{s}$. Several minima are observed. The selection of various lengths
$\ell/2$, $\ell/3$, $\ell/4$,... is thus expected from the above shape of
$E(\xi)$. The more stable local minima is for $\ell/2$, i.e. it corresponds
to ripples of size $\ell$.

For $\xi>\ell/2$, a large potential barrier separates the ripple state from
the ``giant dune" state. The asymmetry of this barrier implies the
irreversibility of the ``ripple $\rightarrow$ dune" process. The ripple
metastability is thus also obtained from a mean-field treatment in
agreement with the above numerical results.

Nevertheless, the mean-field solution does not predict any
disordered dynamical regime. Indeed, fluctuations of $\xi$ are neglected.
Fluctuations destroy the order when $K_s$ is small.

\section{Mapping onto  a two-dimensional
$\,\ell\times\infty\,$ static problem}

Although ripple formation is a {\it dynamical} phenomenon, our study is
generated via Monte Carlo (equilibrium-like) simulations. The dynamical
aspect survives only because the relaxation towards equilibrium is slow
and hampered by metastability.

Another remark concerns Eq.(1). Upon setting
$\delta_{\sigma_i,+1}=\frac{1}{2}(1-\sigma_i)$ in Eq.(1), multiplying out
the $\sigma$'s, and using the fact that $m=\sum_i\sigma_i$ one finds
\begin{equation}
E=-K_{nn}\sum_i\sigma_i\sigma_{i+1}-K_s\sum_i\sigma_i\sigma_{i+\ell}
+\frac{1}{2}K_sm
\end{equation}
in which the last term is an irrelevant constant that is
set equal to zero later. Replace now each site index $i$ by two integer
coordinates $j$ and $k$ uniquely defined by the conditions
\begin{equation}
i\equiv j\ell+k \quad\mbox{ with }\quad k\in\{0,1,\ldots,\ell-1\}
\end{equation}
and write $\sigma_i=\sigma_{j,k}$. In terms of these new variables the
Hamiltonian Eq.(4) takes the form
\begin{equation}
E=-K_{nn}\sum_{j=-\infty}^{\infty}\sum_{k=0}^{\ell-1}\sigma_{j,k}\sigma_{j,k+1}
-K_s\sum_{j=-\infty}^{\infty}\sum_{k=0}^{\ell-1}\sigma_{j,k}\sigma_{j
+1,k}
\end{equation} provided we interpret $(j,k)$ as a site on an
$\ell\times\infty$ cylindrical lattice with screw boundary conditions
(CL/sbc) around the cylinder. Hence $E$ of Eq.(6) describes
Onsager's anisotropic nearest-neighbor Ising model on a CL/sbc, but with
the additional constraint $m=0$.

This sheds an interesting light on the simulation results. The relevant
facts are

(a) The constraint $m=0$ and the sbc
are irrelevant for the thermodynamics of this
model, which is therefore the same as that of the Onsager model
without constraint and in magnetic field zero.\\

(b) The $\,\ell\times\infty\,$ lattice has no phase transition.
But in the limit $\ell\to\infty$ this model does have the
ferromagnetic
Onsager phase transition.
Of course for $\ell$ sufficiently large simulations will show
an order/disorder transition-like behavior.\\

(c) The ground state ${\cal G}_0$,
due to the condition $m=0$, necessarily has an
interface. If the $j$ direction (the one parallel to the cylinder
axis)
is infinite, ${\cal G}_0$ is a cylinder
which is positively magnetized for $j>0$ and negatively magnetized
for $j\leq 0$. If the $j$ direction
is finite (let's say $-J<j\leq J$ for some
$J$), then ${\cal G}_0$ is still
a low-lying state of energy $E({\cal
G}_0)\sim -\ell K_{s}$, but another low-lying state
appears {\it viz.} the state ${\cal G}_1$
which has positive magnetization for $\ell/2\leq k\leq\ell-1$ and
negative
magnetization for $0\leq k<\ell/2$ ($\ell$ supposed even for
convenience). Its energy is $E({\cal G}_1)\sim-2J K_{nn}$.
Both states are local energy minima in configuration space
(because in both the interface length is at a local minimum),
and, with the Kawasaki dynamics of this paper,
the higher one of the two states is metastable
with respect to the lower one.
It is evident that ${\cal G}_0$ corresponds to the "dune" and ${\cal
G}_1$ to the "rippled state".

\section{Connexion to recent underwater experiments}

The physics underlying both ripple and dune formation is far from being
understood making any analytic result valuable. Indeed, it has been
believed that dune dynamics are of a different nature than ripple ones.
Moreover the question of ripple growing to eventually produce a dune is
still open.

Bagnold \cite{bagnold} distinguished ripples and ridges. Periodicity of the
former ones was time independent while periodicity of the latter ones was
growing with time. The present work suggests in a simple way that both
patterns are two aspects of the same phenomenon since the ripple state
seems to be metastable.

Recent experiments of underwater sand \cite{betat,stegner} have exhibited
the formation of ripples. There, sudden merging of ripples do also occur
in qualitative agreement with our results (see the upper part of the
dynamical phase diagram of Figure 4). Moreover, these authors have found
that a ``transition" takes place between a flat landscape regime and ripple
formation at a critical value of the Reynolds number Re. This feature is to
be put in parallel to the transition that we have obtained (see Figure 4)
between a disordered regime and a ripple dynamical regime. Indeed, the
saltation length $\ell$ could be directly connected to the physical
parameter Re.

\section{Comparaison with a cellular automaton}

In addition, a recent cellular automaton model \cite{ijmpc} did produce a
dynamical ripple state which eventually stabilizes into a giant dune. This
model considers a granular landscape for which surface local fluctuations
are dissipated through both saltation and avalanches \cite{btw}. Saltation
is restricted to grains exposed to the wind while avalanches occurs
everywhere.

The associated dynamics is rather complex. First, ripples appear. Secondly,
they coalesce very slowly into ``giant dunes". Main features of ripples and
dunes formation are recovered within the present spin framework except for
the disordered regime which is absent in \cite{ijmpc}.

Eventhough the cellular automaton model as well as actual spin-like model
are far from real desert dunes and ripples, they seems to reproduce some
analogous features.

\section{Conclusion}

In summary, we have proposed a 1-$d$ Ising model for ripple formation. The
spin model has been studied numerically. In addition, a mean-field solution
for the ripple size was also obtained. Moreover, the ripple state has been
found to be metastable. A result which is in full agreement with very
recent experiments under water as well as a previous landscape simulations.

Nature is obviously much more complex that a one-dimensional spin model.
However our very simple description reproduces some basic features of dune
formation as seen above. Moreover, the model could be extended to
include, for instance, a distribution of saltation lengths or even,
different grain types. This is left for future work. We
do hope our results will stimulate additional experimental work
in particular to check our dynamical phase diagram.

\section{Acknowledgments}

NV thanks the FNRS (Brussels, Belgium). He also thanks the LMDH for its
hospitality during the progress of this work. We would like to thank Dr H. J. Hilhorst
for a very stimulating comment.

\newpage

{\noindent \large Figure Captions} \vskip 1.0cm

{\noindent \bf Figure 1} --- (top) Sketch of the rule. A grain at position
$i$ is unstable and is carried over by the wind to the position $i+\ell$
where an avalanche relaxes the surface. Spin signs indicate the slope of
the landscape. (bottom) After the saltation-avalanche event, the
magnetization remains unchanged.

\vskip 1.0cm {\noindent \bf Figure 2} --- Typical space-time patterns in
which domain walls are drawn: $L=144$ and $\ell=21$. Three different sets
of parameter values are shown: (a) $K_{nn}=0.5$ and $K_s=0.5$, (b)
$K_{nn}=2.0$ and $K_s=2.5$, and (c) $K_{nn}=2.5$ and $K_s=0.5$.

\vskip 1.0cm {\noindent \bf Figure 3} --- The density of domain walls
$\rho$ as a function of time. Three different sets of parameter values are
shown, from top to bottom: $K_{nn}=0.5$ and $K_s=0.5$, $K_{nn}=2.0$ and
$K_s=2.5$, and $K_{nn}=2.5$ and $K_s=0.5$.

\vskip 1.0cm {\noindent \bf Figure 4} --- The dynamical phase diagram
$(K_{nn},K_s)$ of the model. The disorder phase is represented in grey in
the diagram. The order phase is represented in white. Ripples and dunes are
also distinguished.

\vskip 1.0cm {\noindent \bf Figure 5} --- The density of domain walls
$\rho$ as a function of the Ising interaction $K_{nn}$. Three different
values of $K_s$ are shown. The crossover separating disorder and order
regimes is emphasized.

\vskip 1.0cm {\noindent \bf Figure 6} --- Typical shape of the energy per
lattice site $E/L$ as a function of the normalized domain size $\xi / \ell$.

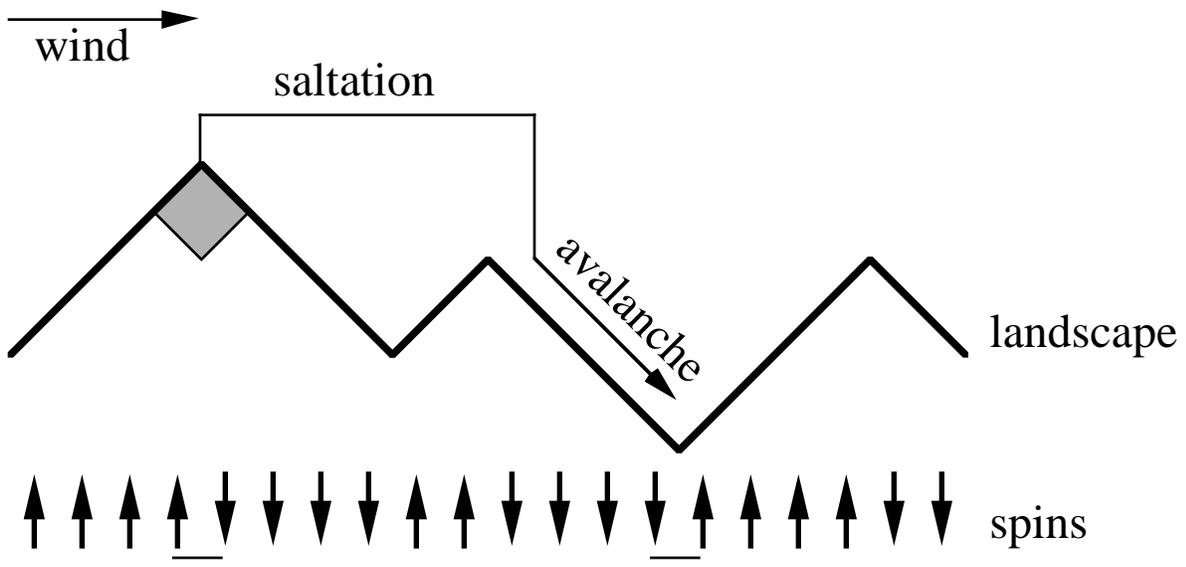
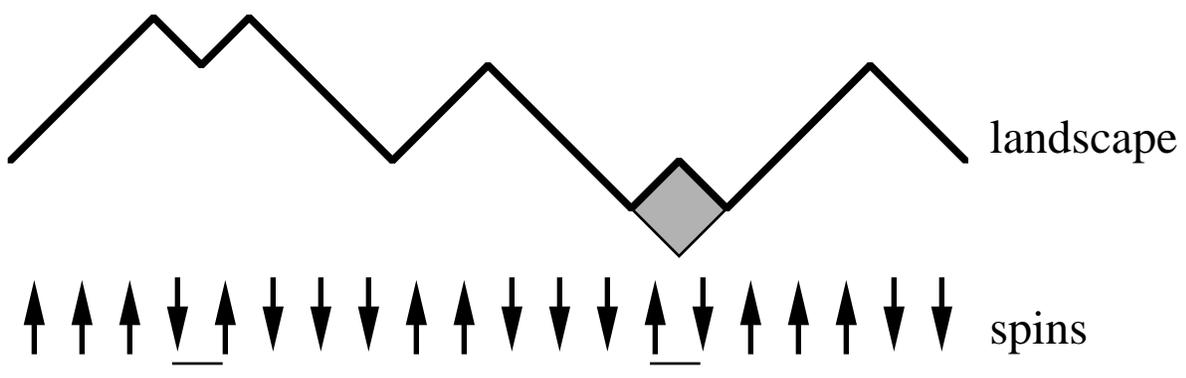
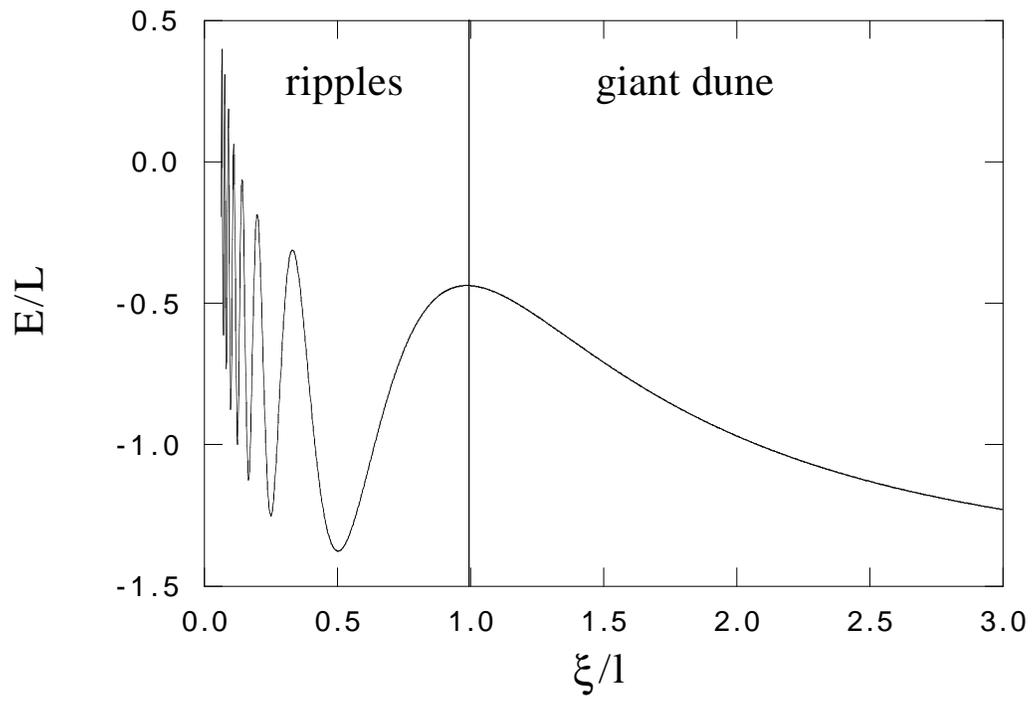
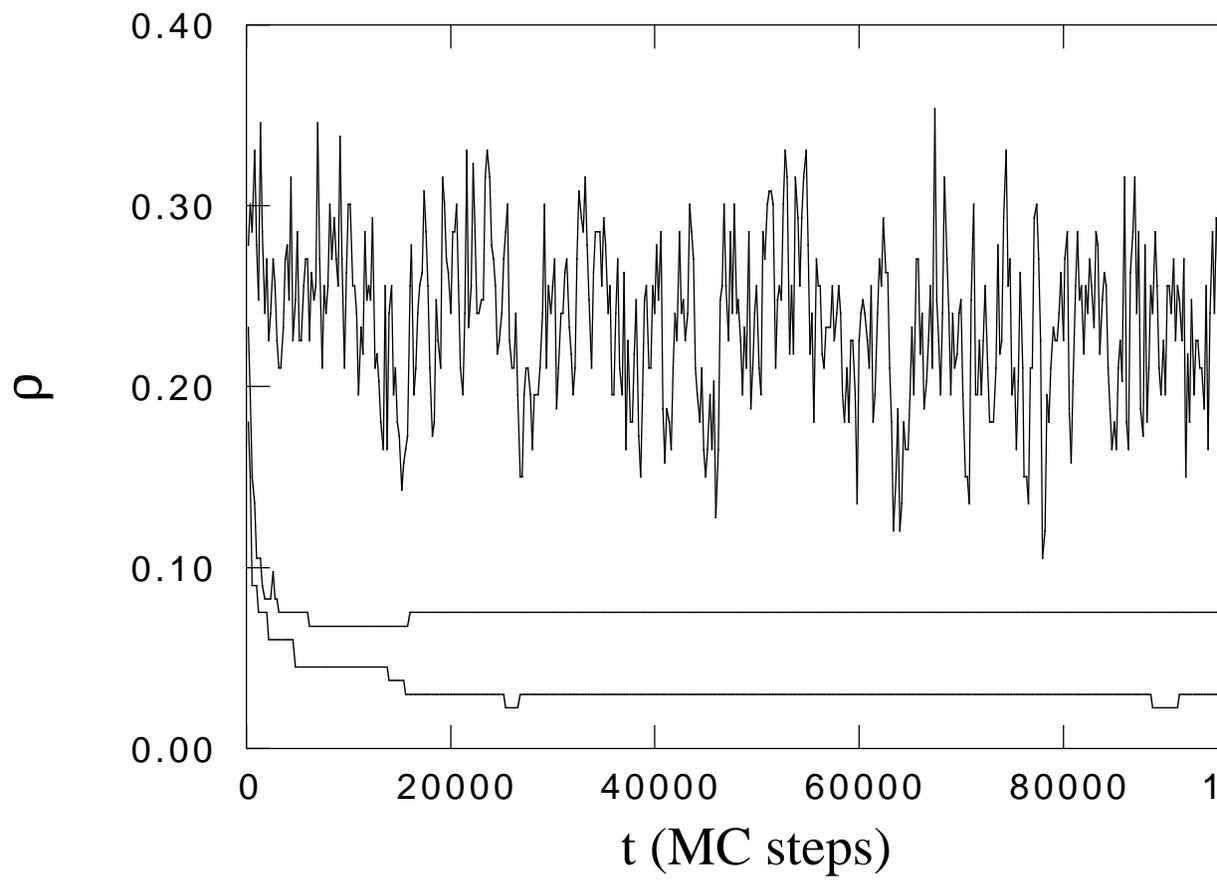
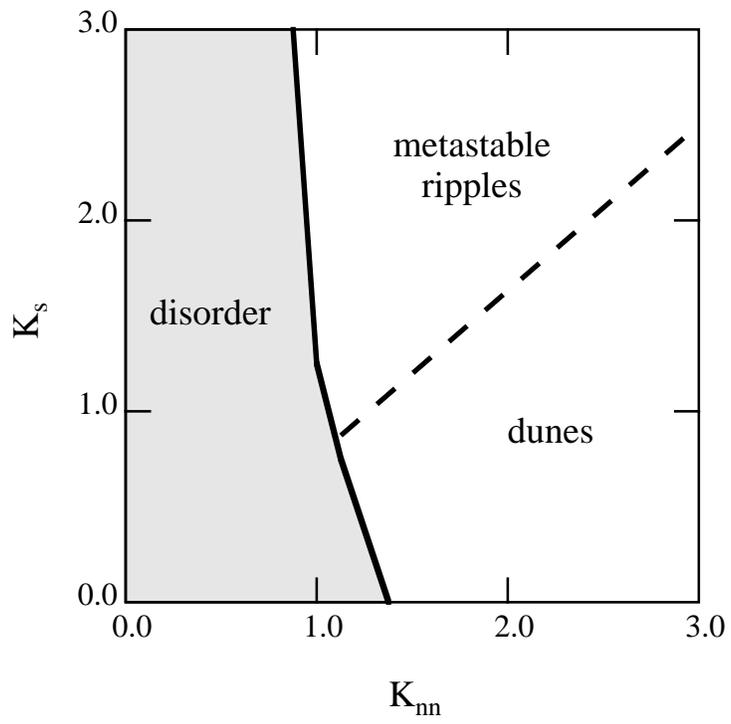
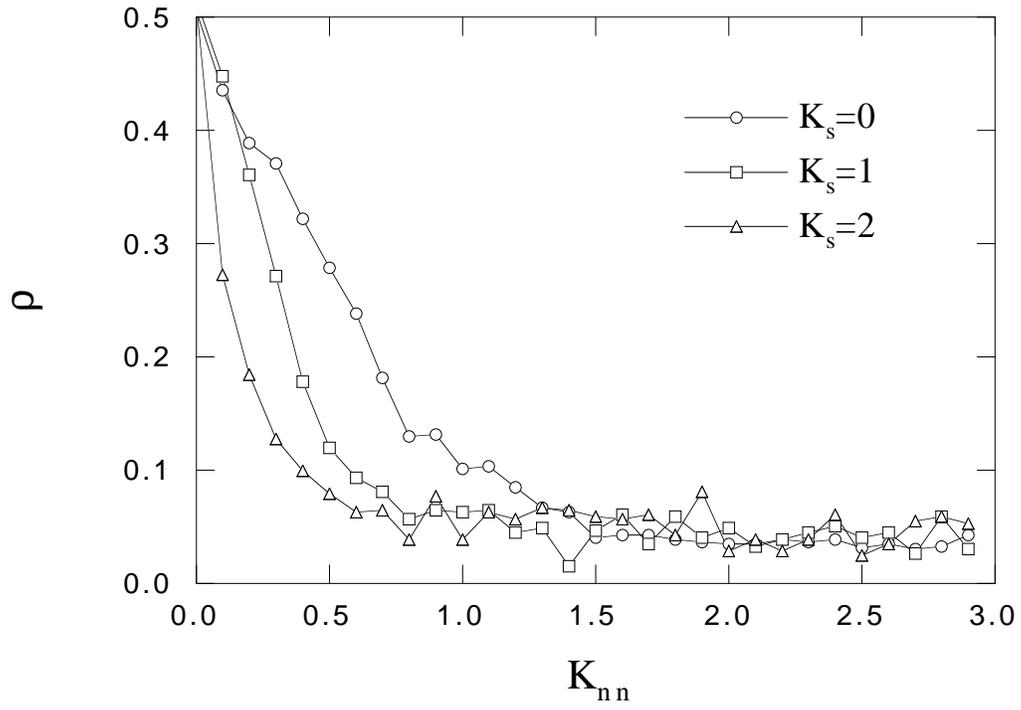
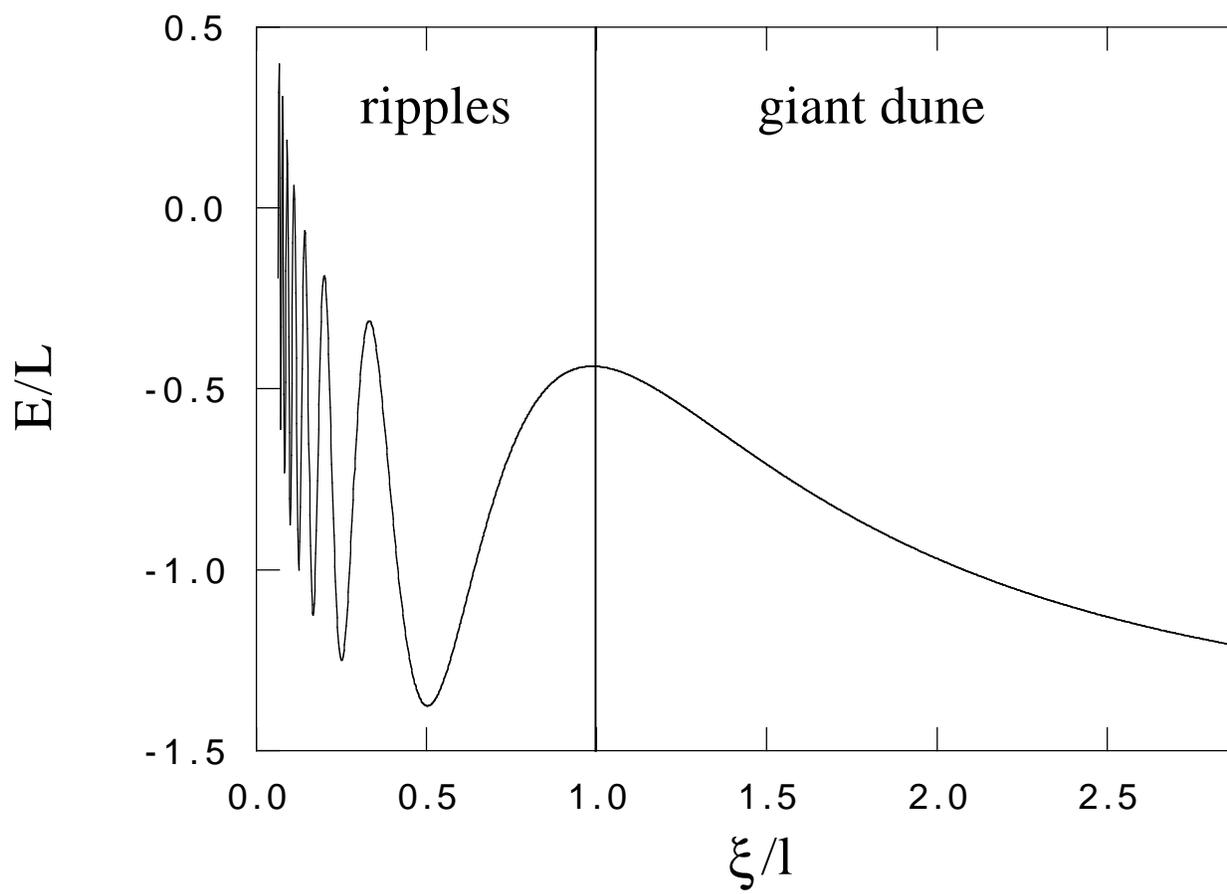
\end{document}